\documentstyle[12pt]{article}

\begin{document}

\begin{center}
{\Large{\bf{ Mott Transition in the Hubbard Model\footnote{ Based on article in
{\bf Frontiers in Solid State Sciences} editors L.C.Gupta and M.S.Multani,
(World Scientific Publications, Singapore).}}}}
\vglue 1.5cm
{\large{\bf B. Sriram Shastry}}\\
AT \& T Bell Laboratories\\
Murray Hill, N.J.,07974\\
\vglue .5cm
[18 August 1992]
\end{center}
\vglue 1.5cm
\begin{center}
{\bf ABSTRACT}
\end{center}
{\small
In this article, I discuss  W.Kohn's criterion for a metal insulator
transition, within the framework of a one band Hubbard model. This
and related ideas are applied to 1-dimensional Hubbard systems, and some
intersting miscellaneous results discussed. The Jordan Wigner transformation
converting the two species of fermions to two species of hardcore bosons
is performed in detail, and the ``extra phases''
 arising from odd-even effects
are explicitly derived. Bosons are shown to prefer zero flux (i.e.
diamagnetism),
and the corresponding ``happy fluxes'' for the fermions identified.
A curious result following from the interplay between orbital
diamagnetism
and spin polarization is highlighted.
 A ``spin-statistics'' like theorem, showing
that the anticommutation relations between fermions of opposite spin are
crucial to obtain the SU(2) invariance is pointed out.
}

\vglue 1cm

\newcommand\cond{\sigma _{xx}(\omega)}
\newcommand\curr{|<0|j_x |\nu>|^2}
\newcommand\ddd{\triangle}
\newcommand\fplus{\Phi_{\uparrow}}
\newcommand\fmin{\Phi_{\downarrow}}
\newcommand\beq{\begin{equation}}
\newcommand\eeq{\end{equation}}
\newcommand\om{\omega}
\newcommand\nh{\hat{N}}
\newcommand{\rref}[1]{{$^{\cite{#1}}$}}
\newcommand\bo{ \beta \omega /2}
\newcommand{\sai}{S^{\alpha}_i}
\newcommand\sbi{S^{\beta}_i}
\newcommand\sgi{S^{\gamma}_i}
\newcommand\chik{\chi_{i,k}}

\newcommand\sak{S^{\alpha}_k}
\newcommand\saj{S^{\alpha}_j}
\newcommand\sbj{S^{\beta}_j}
\newcommand\sgj{S^{\gamma}_j}
\newcommand\chij{\chi_{i,j}}
\newcommand{\al}{\alpha}
\newcommand{\be}{\beta}
\newcommand{\ga}{\gamma}
\newcommand{\la}{\lambda}
\newcommand{\Sy}{ {\cal S}}
\newcommand{\An}{{\cal A}}

\newcommand\dagg{{\dag}}
%\newpage

\section{Introduction}

\qquad In this article, I will give a brief introduction to the Mott
Transition in the Hubbard model. This subject is in a state
of considerable flux, since it is believed by a large number of theorists,
to be at the heart of the High $T_c$ phenomenon. This point of view was
suggested by Anderson\rref{pwa} as early as in 1986, at the Bangalore
ICVF conference. It is less clear as the years go by, that the problem
has an easy solution. However  the validity of the Mott-Hubbard point of view
as a starting point, in the spirit of providing at least the analog of
$H_0$ seems very probable to me,
 of course it may well be that there is a mystery $H_1$
remaining to be discovered! The scope of this article is much more limited.
I would like to  describe the transition, and discuss some
powerful techniques
that have been recently brought to bear on this problem, within a well
defined problem, the 1-d Hubbard model. These ideas have also been applied
with considerable success in higher dimensions, by clever numerical work,
and I will direct attention to these articles at the end.

\qquad In Section 2 we survey the problem of the Mott transition in various
Hubbard like models. In Section 3,  I discuss the criterion of Kohn for the
Mott transition, and define the charge stiffness from the Kubo formula point of
view as well as the energy point of view. In Section 4, I carry out the Jordan
Wigner   transformation and identify extra phases arising from it.  The
diamagnetic inequality for bosons is stated, and an amusing spin statistics
like theorem for  two component hard core boson/fermion system remarked on.
Amusing
odd even effects in diamagnetism and weak spontaneous magnetism in the Hubbard
model are pointed out.  In Section 5, I discuss the Bethe Ansatz equations
for the 1-d Hubbard model, which turn out to be very useful in extracting the
charge stiffness.

\section{The Mott Transition}

\qquad We will first of all dispose of the  long ranged Coulomb
interaction by  confining ourselves to the single band
Hubbard model
: there is no good reason to do this beyond achieving simplicity.
The conducting state, by definition,
 has metallic
screening, making the neglect of the long ranged coulomb interaction a
reasonable approximation. The nature of the insulating state obtained within a
purely short ranged Hubbard model is surely incorrect with regard to the
excitations involving promotion to the upper Hubbard band; i.e.
optical excitations.

\qquad The situation in say the  2-dimensional square lattice is complicated
by magnetic LRO. At half filling, i.e. one electron per site, the ground state
is presumably magnetically ordered {\bf and} insulating. The usual
statement of a Mott-Hubbard gap is in terms of an energy gap towards adding
an extra particle to a half filled band: the existence of a gap of O(1)
distinguishes the insulator from a metal. This formulation, applied to
the magnetically ordered Antiferromagnet, is confused by the existence
of a Spin Density Wave gap, and the subtlety of the Mott-Hubbard gap
is swamped by the essentially trivial SDW gap. We need to consider situations
where the band structure inhibits the formation of an SDW state: many
possibilities arise (a)  The triangular lattice: here the band structure
does not have a strong maximum at any $Q$ at half filling \rref{hrk}
although a three sublattice ordered AFM state is the ultimate large $U$
fate of the model;
(b) Random hoppings: this is a doubly hard problem with disorder
and interactions  and no really convincing studies exist so far;
(c) The Hubbard model on a special   lattice in 2-dimensions constructed
by Shastry and Sutherland\rref{ss}: (see Figure 1). This is an interesting two
dimensional lattice with
a unit cell consisting of four squares of the square lattice
 with diagonal bonds (existing in only two of these squares)
running NE-SW in one square and NW-SE in the
square which shares no bonds with the first. We consider a Hubbard model on
this
lattice with hoppings $t$ and $t \sqrt{2 \alpha}$.
The band structure gives
a semi-metal at half filling, with a quadratic touching of the filled
valence band, and an empty conduction band, for generic
$\alpha$. At large $U$, the model reduces
to a Heisenberg Antiferromagnetic model with $S=1/2$ on this lattice,
with bond strengths say $J= 4t^2/U$ and $J'=2 J \alpha$; the really interesting
fact is that the
Heisenberg model can be solved exactly, and has frozen dimer order for
the region
$\alpha \ge 1$. For
$\alpha <1$ the model has not been solved, and approaches the usual nearest
neighbour model as $\alpha \rightarrow 0$.
This is the only case of an exactly solvable Heisenberg model
with $S=1/2$  in
2-dimensions that I am aware of, and is of interest from a Mott-Hubbard
point of view, since one expects the semi-metal to insulator transition, to
be quite representative of the phenomenon, uncomplicated by magnetic LRO.

\qquad We confine ourselves in   this article
to the 1-d Hubbard model,
which has the great virtue of being solvable exactly  by the Bethe Ansatz
\rref{lieb}, and also of exhibiting a Mott Transition, uncomplicated
by magnetic LRO. The model turns out to possess algebraically
decaying spin correlations, which are by uncertainty principle arguments,
the maximal remnants of magnetic LRO in 1-dimension.
The tool that we bring to bear on this problem, is a powerful
and beautiful
argument due to W.Kohn\rref{kohn}. This argument of Kohn,
recently sharpened and revived by us \rref{sskohn}
after many decades of neglect,  turns out to be quantitatively
implementable in the 1-d Hubbard model. In essence, Kohn's argument is
that we can distinguish between a metal and an insulator, and hence track
the Mott transition, by studying the ground state energy as a function of
the boundary angle. This involves studying the solution of
the many body problem with twisted boundary conditions, and happily,
the Bethe Ansatz method admits this complications and remains tractable
\rref{sskohn}.

\section{Kohn's criterion for the Mott transition}

\qquad We first review the arguments of Kohn, specializing
to a 1-band d-dimensional lattice fermi system, and deduce the
implications- which concern the utility
of studying the effect of twisting the boundary conditions
in a many body system. The second derivative of the ground state energy
under variation of the flux is related to the zero frequency
physics of the electrical conductivity. The economy of the Kohn argument
is noteworthy: we talk of the properties of the  conductivity directly,
and do not worry about
 one electron gaps. It should be mentioned that several works have appeared in
literature,
trying to use various indirect probes of the Mott transition, including
the double occupation, binding between holes and doubles, and various
fictitious
fermi-bose fields. The directness of the Kohn argument,
in contrast, is particularly
appealing.

Consider a d-dimensional hypercubic lattice
 of linear dimension L, with
spinless fermions  having a
nearest neighbour hopping matrix element `t',
and with arbitrary density dependent interactions that are
lattice translation invariant and assume periodic bc's.
The case of particles with spin is trivially obtained from these
considerations, in general the different species can be twisted in
differing degrees, leading to charge as well as a spin stiffness.
We now
introduce a uniform vector potential $A_x \hat x$, which modifies
the hopping in $\hat x$ directed bonds by the usual Peierls
phase factor $t \to t\, exp( \pm i \Phi /L)$, where
$\Phi =L( A_x e/\hbar c)$ and the lattice constant $a_0=1$.
Expanding
the exponential out we find the perturbed Hamiltonian
$H'=H - \Phi j_x/L - \frac{1}{2} \Phi ^2 T_x /L^2 +O(\Phi ^3)$,
where $j_x= 2t \sum sink_x C_k^{\dagger}C_k$ and
$T_x=-2t \sum cosk_x C_k^{\dagger}C_k$, and $H$
is the Hamiltonian for the interacting fermi system.
The energy shift of the ground state (g.s.)
in the presence of the field
is $E_0(\Phi)-E_0(0)\equiv D \Phi ^2/L^{2-d}
+O(\Phi^4)$, or equivalently, we  define the charge stiffness D:
\beq
D=(\frac{L^{2-d}}{2}) \frac{d^2 E_0(\Phi)}{d \Phi^2} /_{\Phi=0}.
\eeq
 The stiffness constant D is given by
second order perturbation theory as
\beq
 D= \frac{1}{L^d} [\frac{1}{2d} <-T> -
\sum_{\nu \ne 0} \frac{\curr}{E_\nu -E_0
}] \label{stiffness}
\eeq
where $<T>$ is the kinetic energy expectation in the
g.s. and $E_0(0)\equiv E_0$.
 We have assumed that $<j_x>$ is zero, this is not always true, and some
examples are given in Sec(4).
Higher order  (non quadratic) terms
in the energy shift formula
are important when the
energy shift is comparable to the energy gaps in the spectrum
of $H$. The
latter are $O(1/L)$ in metals and so in this case
corrections arise when  $\Phi$
is  of $O(1/L^{(d-1)/2})$.
Level crossings would occur and perturbation theory
would breakdown for $\Phi$  of $ O(\pi)$. The region in
$\Phi$ wherein the derivatives are unambiguously defined,
therefore shrinks with increasing L in two and higher dimensions.

\qquad  We  next specialize to $A_x \to A_x^0 exp
(-i \omega t)$  leading to an electric field $\vec E_x=A_x(i
\omega/c)
\hat x$,
 from which the usual linear response formula gives the
imaginary part of the
ac conductivity
\beq
 \Im m \; \cond = \frac{2e^2}{L^d \hbar ^2 \omega}
[\frac{1}{2d} <-T> - \wp \sum_{\nu \ne 0}
\frac{\curr (E _{\nu}-E_0)}{(E_{\nu}-E_0)^2-\hbar ^2\omega ^2}
]. \label{imcond}
\eeq
{}From Eqs(\ref{stiffness}) and Eq(\ref{imcond})
 we see that $\lim_{\omega\to 0} \omega \;
\Im m \;\cond =
\frac{2 e^2}{\hbar ^2} D $, and  also
 $\lim_{\omega\to \infty} \omega \;\Im m\;\cond =
\frac{ e^2}{ d \hbar ^2 L^d} <-T>$.
The high frequency behaviour of the imaginary part
of the conductivity
implies for the real part, through the usual
 dispersion relations, the well known\rref{fsumrule}
$f$-sumrule:
\beq
\int_{-\infty} ^\infty
 \Re e\; \cond \,d \omega = \frac{\pi e^2}{d \hbar^2 L^d}<-T>.
\eeq
 More interesting is the small $\omega$
behaviour. From standard dispersion theory we separate the low frequency part
and conclude  that
\beq
\Re e \;\cond = \frac{2 \pi e^2}{\hbar}[
 D \delta (\hbar  \omega) +\frac{1}{L^d} \sum_{\nu \ne 0}
\curr \delta ((E_{\nu}-E_0)^2-\hbar ^2\omega ^2 )] \label{recond}
\eeq
D, the coefficient of $\delta (\hbar \omega)$ , if non-zero,
implies free acceleration  or infinite d.c. conductivity,
which is reasonable here since there is no
dissipative mechanism in the model at $T=0^0$.
The coefficient D is  essentially the inverse
of the effective current carrying mass (for free electrons it
is $\pi \rho e^2/m$). Therefore the $f$-sumrule
is satisfied by the {\it sum of
two terms of the same order}, the stiffness $D$,  and the
 ``intraband dipole matrix elements''.
I would like to remark that
Eq(\ref{recond}) was not written down by Kohn,
instead he   emphasized the imaginary
part of the conductivity (i.e. Eq(\ref{imcond})), and infact discussed metals
as systems for which the $f$-sumrule is violated. A slight
change of view point  is advocated in the Shastry-Sutherland
paper \rref{sskohn}, wherein the real part of conductivity, Eq(\ref{recond})
is explicitly stated, and it is stressed that there is no violation of
the $f$-sumrule, the zero frequency piece, together with the finite frequency
piece certainly fulfil it. This point of view makes the charge stiffness D
to appear very much like the stiffness in a superfluid or a superconductor.

\qquad   This is all very nice, but we still need an algorithm to calculate
the stiffness! Here Kohn makes the  important observation that
changing boundary conditions is equivalent to applying the $\Phi$ field, we
can absorb the Peierls phases by a pseudo gauge-
 transformation\rref{kohn,yang}
and shift the effect of $\Phi$ into twisted bc's for the
wavefunctions :
\beq
\Psi (\ldots \vec r+ L \vec x \ldots)= exp\{i\Phi\}
\Psi (\ldots \vec r \ldots) .
\eeq
Hence, a calculation of the ground state energy with twisted b.c's
suffices to give D.

\qquad A crucial point (familiar from Landau's fermi liquid theory)
is that for a Galilean invariant interacting system,
an analogous calculation would give the coefficient of
$\delta(\hbar \omega)$  in Eq(\ref{recond})  unrenormalized  by interactions
since $[j_x,H]=0$ ( the first term in Eq(\ref{recond}) becomes the particle
density).
For lattice fermions the operator $j_x$ commutes with the
hopping part of H, but not with the interaction piece in
general
and hence for {\it interacting lattice fermions} there is the
possibility that the two terms in Eq(\ref{stiffness}) cancel as some parameter
is varied,  signalling a metal insulator transition. The absence
of Galilean invariance, thus allows the charge carrying effective
mass to vary with interactions,  and in fact  to diverge.

\qquad We thus see that Kohn's argument is a means of obtaining
a ``transport coefficient'' D, from a calculation of the ground state
energy. This is less uncomfortable when we recognize that D is {\it not}
a characteristic of energy dissipation, but rather measures the quantity
of freely responsive fluid. In fact it is obvious that the stiffness
so defined is precisely the superfluid stiffness in say a superconductor.
Thus Kohn's metal is a superconductor as well! True superconductivity
may be distinguished from the Mott-Hubbard-Kohn metal by introducing disorder.
True superconductors are insensitive to disorder by the Anderson theorem,
whereas the M-H-K metal would become a dirty metal, with a finite
 resistivity, and with energy dissipation. This brings us to the work of
 Thouless\rref{thouless}, who brought a closely related
and  successful idea to the case
of dirty metals, to distinguish between a dirty metal and an insulator,
i.e. to study the Anderson transition. Thouless's ideas are worked out in a
purely
non-interacting limit, where one examines the various
one-electron eigenstates of a random potential. The conducting states
have a finite curvature with boundary angle, but the curvatures of successive
energy levels are typically opposite in sign (since the dirty metal is
crudely visualizable as a ``periodic'' system with a very short period in
k-space
), whereby a sum over the energy levels would give a thermodynamically
negligible stiffness. The insulating states (i.e. localized states)
have a vanishing  stiffness, typically exponential in system size with
 a localization length setting the scale.  A  highly nontrivial part of
the Thouless argument concerns the disorder average of the stiffness
for the extended states, and relating this to a truly dissipative object,
the conductance. This relationship has been  explicitized in a recent
work by Akkermans and Montamboux\rref{akker}.

\qquad   It is worth noting that the relationship between D and
the ground state energy (g.s.e.) is generalizable to finite temperature, with
the
free energy  $F(\Phi)$ replacing the g.s.e in Eq(1). This relationship does not
seem particularly
useful for the Mott problem, since at finite temperatures, D is expected
to vanish exponentially with system size since there exists a thermal length
scale  causing an exponential decay of the one particle correlation
function.

\qquad I would like to mention a beautiful (if somewhat inaccessible )
article on superfluidity
by Leggett\rref{ajl} in 1973, which clarifies the role of twisting boundary
conditions in normal fluids, solids and superfluids within the context of
rotating bucket like
experiments, and illustrates the correspondence between these systems and
between metals,
insulators and superconductors. Also the work of Fisher et. al.\rref{fisher}
makes  precise the connection between the helicity modulus (i.e. $d^2F(\Phi)/
d\Phi^2$) and the superfluid stiffness. A recent paper by Scalapino, White and
Zhang \rref{scalzhang} discusses the various limits $\vec q \rightarrow 0$ and
$\omega \rightarrow 0$, for the electromagnetic response
kernel in 2 and higher dimensions, with the object of clarifying the role
of the shrinking region of $\Phi$ in the Kohn argument alluded to above.
\section{1-d Hubbard model: Generalities and Curiosities}
\qquad In this section we specialize to the 1-d Hubbard model, which was solved
by Lieb and Wu \rref{lieb} in 1968. Before discussing the Bethe Ansatz method,
we discuss some general ideas concerning the stiffness, and the problem of
particles moving in a flux.
We pay particular
attention to the Jordan Wigner transformation in this problem, which introduces
a certain  ``flux'' of its own, leading to occasionally surprising results.
It is convenient, and natural, to convert the model to one involving hard-core
bosons.
Let us recall the Jordan Wigner transformation between Pauli matrices
$\sigma_n^\al$, with $1 \leq n \leq L$ and fermions $c_n$:
\beq
\sigma_1^z \ldots \sigma_{n-1}^z \sigma_n^{+}= c_n^{\dagger}, \label{jw}
\eeq
with $\sigma^z_n=2 c^{\dagger}_n c_n-1$, and the hermitean conjugate of the
Eq(\ref{jw}). It is straightforward to see that
\beq
c^{\dagger}_n c_{n+1}= - \sigma^{+}_n \sigma^{-}_{n+1}
\eeq
and
\beq
c^{\dagger}_L c_{1}= - (-1)^{\nh -L+1}\sigma^{+}_{L} \sigma^{-}_{1}.
\eeq
where $\nh= \sum_{i}c^{\dagger}_ic_i$, is the number operator.

\qquad We are interested in converting a fermionic hopping hamiltonian having
density dependent interactions, and
with a
boundary angle (or flux) $\Phi_f$, and with $c_{L+1}=c_1$,
\beq
H_f= -\sum_{n=1}^L (c^{\dagger}_n c_{n+1} \exp(i \Phi_f/L)+h.c.)+ V_{int}
\eeq
 to a hardcore boson hopping problem with
a boundary angle $\Phi_b$, and $\sigma_{L+1}=\sigma_1$,
\beq
H_b= -\sum_{n=1}^L (\sigma^{+}_n \sigma_{n+1} \exp(i \Phi_b/L)+h.c.)+ V_{int}.
\label{boson}
\eeq
 We can do this in two stages, firstly use the Jordan-Wigner transformation
Eq(\ref{jw}), and next a unitary transformation $U_1=\sum_n \exp(i n \chi/L)$,
with an appropriate $\chi$, together this yields
\beq
\Phi_f=\Phi_b+ \pi (\nh-1); \; \mbox{ mod 2$\pi$} \label{fermibose}.
\eeq
The reason for preferring to work with bosons is that there are no extra phases
involved. In fact we can show easily that the boson problem
Eq(\ref{boson}), generalized to
any dimensions, and on any lattice, has a preference
to having zero flux, i.e.
\beq
E_{b}^0(\Phi) \geq E_{b}^0(0), \label{ineq}
\eeq
where $E_{b}^0(\Phi)$ is the ground state energy in the presence of the
flux $\Phi$. This  {\bf diamagnetic inequality} follows from writing the
eigenvalue
equation for Eq(\ref{boson}) in a standard basis, with particles
 living  at  various sites, whence the phases make their appearance in the
off diagonal elements of the matrix. We can decrease the ground state energy
by increasing the absolute value of the off-diagonal matrix elements by virtue
of the Rayleigh-Ritz variational principle. Inequality(\ref{ineq}) follows
on using
$|\cos(\theta)| \leq 1$ . Unless the magnetic
field is a pure gauge field, we must maximize at least one ``bond'',
and hence the Inequality(\ref{ineq}), is in general, a strict one and
 may be regarded as a statement of diamagnetism of bosons. A corresponding
theorem for fermions does not exist. In fact the Eq(\ref{fermibose}) tells us
that given the preference of bosons, the fermions
like  (i.e. the minimizing boundary angle is) to have  ``happy'' phases:
\beq
\Phi_f^*= 0, \; \pi \mbox{ for $\nh$=  odd, even}.   \label{fistar}
\eeq
For the Hubbard model, we need a two component Jordan Wigner transformation,
converting bosons $\sigma^\al_n, \; \tau^\al_n$, to fermions $c_{n,\uparrow},
c_{n,\downarrow}$, through
\beq
\sigma_1^z \ldots \sigma_{n-1}^z \sigma_n^{+}= c_{n,\uparrow}^{\dagger}
\eeq
\beq
\sigma_1^z \ldots \sigma_{L}^z \tau_1^z \ldots \tau_{n-1}^z
 \tau_n^{+}= c_{n,\downarrow}^{\dagger}
\eeq
with  $\sigma^z_n=2 c^{\dagger}_{n,\uparrow} c_{n,\uparrow}-1$, and $\tau^z_n=2
c^{\dagger}_{n,\downarrow} c_{n,\downarrow}-1$. The corresponding phase
relations are easily seen to be
\beq
\Phi_{f,\uparrow}=\Phi_{\sigma}+ \pi (\nh_{\uparrow}-1) \label{fermiboseup}.
\eeq
\beq
\Phi_{f,\downarrow}=\Phi_{\tau}+ \pi (\nh_{\downarrow}-1)
\label{fermibosedown}.
\eeq
with obvious notation. We have thus converted the two fermi component
model to one with two hardcore boson species, with given phases.
 It may appear that the phases of the fermions were
completely disposable and not very fundamental. However, at this stage, I would
like
to inject an essentially elementary digression into the discussion,
showing how important the fermi phases really are.

\subsection*{ Digression: Absence of SU(2) for hard core bosons}
\qquad The global SU(2) invariance of the Hubbard model is well known,
and together with the particle hole symmetry, gives the model SO(4) symmetry,
as recently emphasized by Yang and Zhang\rref{yangso4}. This  represents the
(maximal) internal symmetry
of the system. One may cavalierly expect that the hard core bose representation
 also possesses SU(2) symmetry. This is not so. An elementary demonstration of
this follows from the examination of the commutation relations of the
objects:
\beq
T^z=\sum_1^L (\sigma^z_i-\tau^z_i)/2
\eeq
\beq
T^{+} = \sum_1^L (\sigma^{+}_i \tau^{-}_i),
\eeq
and its hermitean conjugate.  The commutator of $T^{\al}$ and $T^{\be}$
{\bf does not} follow the usual SU(2) relations, and nor do $T^{\al}$
commute with the two component boson version of the Hubbard Hamiltonian
. The SU(2) relations of course
follow, if one reinstates the Jordan Wigner strings. {\it This is a kind of
``spin-statistics'' theorem in nonrelativistic physics}.

\qquad Getting back to our main theme,  we examine the implications of our
discussions for the stiffness. The fact that the  minimizing flux
for fermions is $0$ or $\pi$
 implies that we need not necessarily expect a positive definite stiffness D
for fermions, in contrast to the bosonic case. In all cases, we should
examine the behaviour for an even number of particles closely, since here,
the state with flux $\pi$ is the minimum.
The XXZ model,
i.e. the Heisenberg Ising model\rref{yang2},  can be interpreted  as an
interacting one component fermi gas.  When the number of particles
$\nh$ is even,  one has  a non-zero momentum in the ground state,
i.e $<j_x> \neq 0$.
For example think of  2 non interacting particles in a 6-membered ring, we have
a degeneracy with momentum $\pm \pi/3$, and not much can be said about the
stiffness in general terms, since the energy shift has a linear term in $\Phi$.
An explicit calculation using the Bethe Ansatz was given for
the stiffness in this model in \rref{sskohn}, which may be consulted for
details. The model has an interesting Metal-Insulator transition,
with a crystalline solid phase, and the stiffness has a jump discontinuity
at the transition point, very much like the superfluid stiffness jump in the
2-d classical x-y description of a bose fluid.

\qquad In the case of the 1-d Hubbard model, the situation is much richer.
The method of calculating the stiffness will be discussed in the next section,
here one may just assume that  numerical calculations have been performed.
At half filling, i.e. $\nh_{\uparrow}=\nh{\downarrow}=L/2$, an
interesting phenomenon was first
noted by Scalapino and co workers \rref{djs}. They noted that the stiffness D
is infact negative, for $L=4*integer$, but is positive for $L=4*integer+2$.
The stiffness can be calculated for arbitrary system sizes\rref{charles}, and
infact,
vanishes exponentially with system size with a characteristic behaviour
$D=exp(-L/\xi)$, with a known $\xi$, as one would expect for
an insulator. An explanation of the sign was provided \rref{charles}
using the Aharonov Bohm effect of a hole and a double being created,
and transported around the Hubbard ring. A more direct understanding of this
result follows from the considerations above. At $L=4 *integer$, the number of
up and down particles is  even, and hence the bosons are ``unhappy''. The
state at zero flux is higher in energy than that at flux $\pi$, and hence
the stiffness is not necessarily positive. The fact that it is negative, does
not follow
from this ( the energy could, for instance, have a local minimum at flux
$0$), and requires actual calculation. These considerations, turn out
to be not as academic and exotic as the reader may suspect. The negative
stiffness is infact known, and has been observed\rref{aroma} in aromatic
organic compounds, which are
longer analogs of Benzene. The compounds are [16]-annulene and [24]-annulene,
where NMR shows orbital paramagnetism, and these compounds are essentially
the Hubbard model at half filling for L=16,24.

\qquad Another set of curiosities concern the case when the
number of particles is $4*integer$,
away from half filling, in the Hubbard model. At  half filling,
 the ground state is a singlet due to a theorem of Lieb\rref{lieb2}. Away from
half filling, however, whenever the
number of particles is $4*integer$, the ground state is a spin-1 state. Figure
2. shows the singlet and the triplet state energies for a ring of 8 sites,
with 8, 6 and 4 particles. The noteworthy case of 4 particles has a triplet
ground state.
For  $L>8$, similar results hold.
This result comes about due to an interesting and
subtle effect: {\it the spin zero state has orbital frustration, i.e. unhappy
phases in the bosonic representation, which is relieved by partially spin
polarizing the system}. The ground state can be easily shown to have
$|S_{tot}| \leq 1$, by an argument analogous to  that in the Lieb Mattis
theorem\rref{lm}. We can go to a subspace where the $S_z=1$, having an odd
number of
up and down electrons, and in this sector, there are no frustrating phases, and
hence a unique ground state exists. This state cannot be orthogonal to the g.s.
of the non interacting case since the interaction only affects the diagonal
matrix elements; for zero $U$ the ground state is degenerate
between $S=0,1$.
Hence  the spin of the state
could
be zero or unity, but  no higher. For infinite $U$, it is readily shown that
$S=1$, from the Bethe equations which become trivial (see Eq(\ref{bethe})),
whereby one expects $S=1$ for all $U$. The numerics confirm this.  It is
fascinating to speculate that this spin polarization, arising from
orbital paramagnetism, may be a more widely useful concept.

\section{1-d Hubbard: Bethe Equations}
\qquad We next consider the Bethe Ansatz equations, for the 1-d Hubbard model,
in the presence of the flux\rref{sskohn}, generalizing the Lieb Wu equation
\rref{lieb}. Firstly, it is necessary to show that the
Bethe Ansatz goes through in the presence of the flux: this is not obvious
and should be checked. We checked it, and the details are
available\rref{sskohn}.
Basically, the fact that the $R$ matrix in the problem conserves the
nuber of particles, is sufficient to guarantee the Bethe solvability.
The equations are,
\beq
Lk_n=2 \pi I_n+\Phi_{\uparrow}^{f}+ 2\sum_{j=1}^M \arctan[4(\Lambda_j-
\sin(k_n))/U], \label{bethe}
\eeq
and
\beq
2 \sum_{n=1}^N \arctan[4(\Lambda_j-\sin(k_n))/U]= 2\pi
J_j+(\Phi_{\downarrow}^f-
\Phi_{\uparrow}^f)+2 \sum_{i=1}^M \arctan[2(\Lambda_j-\Lambda_i)/U],
\eeq
where L,N,M are respectively the numbers of sites, particles and
downspin particles. The quantum numbers
\beq
I_n=\mbox{integers (1/2 odd integers) : M even(odd)};
\eeq
\beq
J_j=\mbox{integers(1/2 odd integers)  : N $-$ M odd(even)}.
\eeq
The total energy is given by $E_{tot}=-2 \sum \cos (k_n)$.

\qquad The case of $\Phi_{\uparrow}=-\Phi_{\downarrow}$, gives us
the spin-stiffness in the problem, which turns out to be easily computable, and
related to the spin susceptibility, by a fairly rigorous calculation
\rref{sskohn}. The other case $\Phi_{\uparrow}=\Phi_{\downarrow}$,
giving the charge stiffness, is not computable with the same rigor,
since we note that the energy change due to the twist is not extensive, nor is
it O(1), it is infact O(1/L) in total energy. Morover,  subtle level
non-crossing
effects can and do happen \rref{ss3} when the flux is of $O( \pi)$, and one may
not be able to continue a solution found at zero flux to a large value of the
flux.
The usual trick of converting
the sums to integrals does not guarantee the  accuracy of
the result to O(1/L), and the answers obtained by this technique, should be
verified by other means, such as numerical studies.
 Solutions of the integral equations can be  found by
manipulating the integral equations, assuming that they hold for large enough
flux.  The justification of
these manipulations seem to require at least integrability (which we happen to
have for the Hubbard model\rref{bss}), and also the avoidance of unfavourable
fractions in filling.
 The theory of handling these
transcendental equations is not quite complete at this moment,
however the careful work of Woynarovich \rref{woyn} and others has made a
promising start
in this direction. Also  solutions using ideas from conformal invariance
give useful explicit answers \rref{japs}.
The conclusion is that the
stiffness vanishes as we approach half filling linearly in the departure from
half filling. This implies that the Mott transition is directly visible in the
charge stiffness.

\qquad Similar ideas can be applied in say 2-dimensions, and interesting work
of Millis and Coppersmith\rref{millis}, and    Poilblanc and
Dagotto\rref{degatto} shows that the 2-d Hubbard
model, as well as a t-J model show similar signs of Mott transition. Further
applications to a general interacting electron gas have appeared
recently\rref{moul}.
The
utility of these calculations in explaining the optical conductivity
experiments\rref{gordon} in Hi-Tc is noteworthy, and together with the the
$f$-sumrule, the
charge stiffness gives a good insight into the transfer of optical weight
observed, within a one band Hubbard description.

\section*{Acknowledgements}

 I take this opportunity to thank S.C.Zhang, C.Stafford, T.Giamarchi, O.Narayan
and B.Sutherland and E.H.Lieb
for helpful comments and discussions.

\clearpage

\section*{Figure Captions}
\begin{enumerate}
\item The special lattice in two dimensions for which the Heisenberg $s=1/2$
antiferromagnet is exactly solvable.
\item The spectrum of the Hubbard ring with 8 sites. The numbers (N,S)
represent the number of particles and the spin of the state.
\end{enumerate}

\end{document}